\documentclass{egpubl}
\usepackage{pg2022}
 
\SpecialIssuePaper         % uncomment for final version of Computer Graphics Forum, special issue
\usepackage[T1]{fontenc}
\usepackage{dfadobe}  

\usepackage{cite}  % comment out for biblatex with backend=biber
\BibtexOrBiblatex
\electronicVersion
\PrintedOrElectronic
\ifpdf \usepackage[pdftex]{graphicx} \pdfcompresslevel=9
\else \usepackage[dvips]{graphicx} \fi

\usepackage{egweblnk} 
\usepackage{amsmath}
\usepackage{amssymb}
\usepackage{algorithm}
\usepackage{algorithmic}
\usepackage{bbding}
\usepackage{pifont}
\usepackage[labelfont=bf,textfont=it]{caption}

\title{Spatio-temporal Keyframe Control of Traffic Simulation \\ using Coarse-to-Fine Optimization}

\author[Y. Han \& H. Wang \& X. Jin]
{\parbox{\textwidth}
{\centering Yi Han$^{1}$\orcid{0000-0002-9548-7979} \qquad He Wang$^{2}$\orcid{0000-0002-2281-5679} \qquad Xiaogang Jin$^{1}$\thanks{Corresponding author: jin@cad.zju.edu.cn}\orcid{0000-0001-7339-2920}}
\\
\parbox{\textwidth}
{\centering $^1$State Key Lab of CAD\&CG,  Zhejiang University, Hangzhou 310058, China  \\ $^2$University of Leeds, United Kingdom}
}

\begin{document}

% uncomment for using teaser
% \teaser{
%  \includegraphics[width=\linewidth]{eg_new}
%  \centering
%   \caption{New EG Logo}
% \label{fig:teaser}
%}

\teaser{
\setlength{\belowcaptionskip}{0.3cm}
\includegraphics[width=0.63\linewidth]{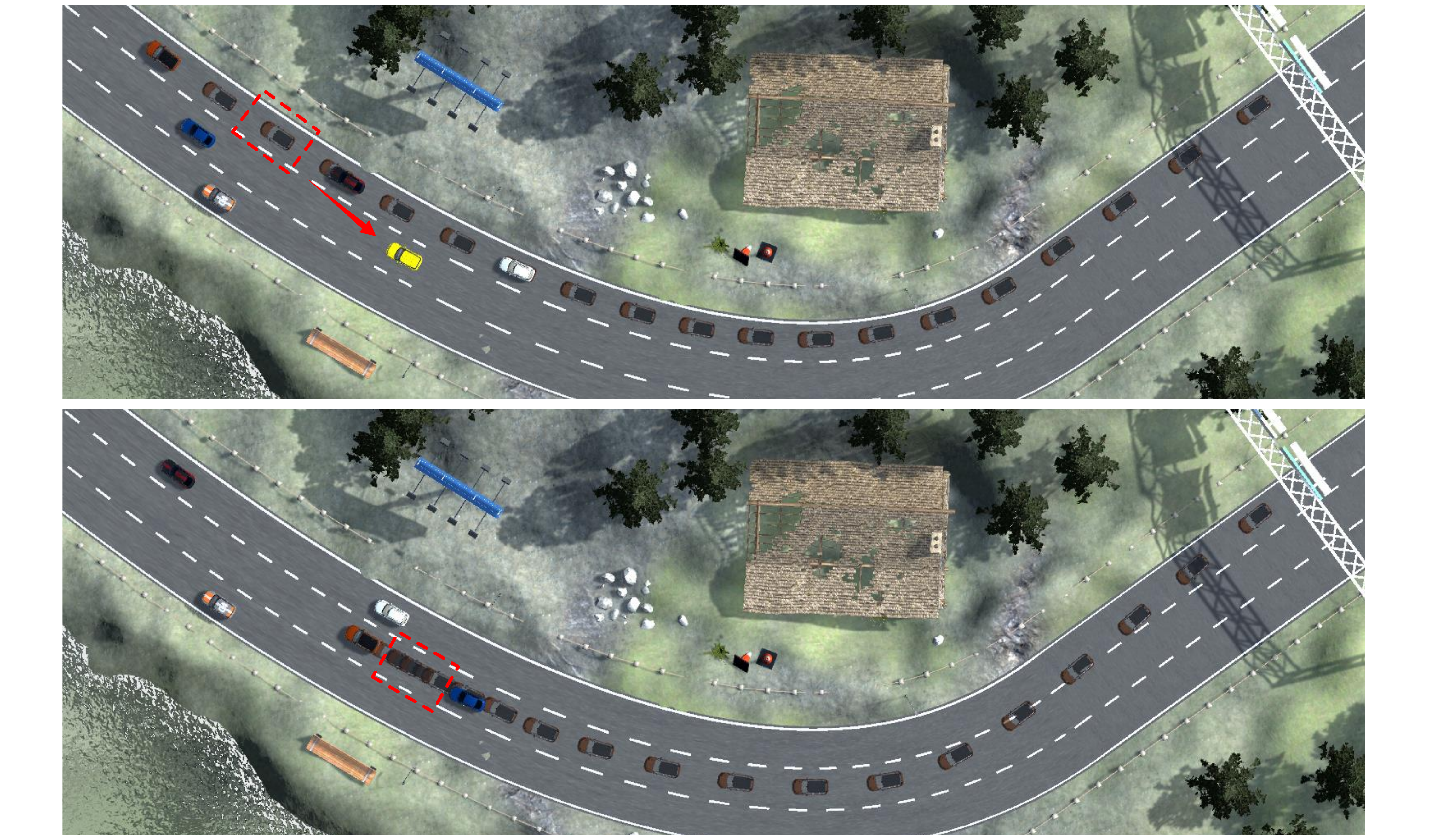}
\centering
\caption{An example of the controlled simulation with a keyframe. (a) The original trajectories. (b) The keyframe controlled trajectories. The vehicle drives along the rightmost lane originally. A keyframe is assigned in the center lane, which denotes that we want the vehicle to arrive at the position marked by the yellow vehicle at the time framed by the red box. As a result, a new reference path is planned through the position at first, and then the vehicle follows it to meet the spatial-temporal constraints smoothly.}

\label{fig:teaser}
}

\maketitle
%-------------------------------------------------------------------------
\begin{abstract}
We present a novel traffic trajectory editing method which uses spatio-temporal keyframes to control vehicles during the simulation to generate desired traffic trajectories. By taking self-motivation, path following and collision avoidance into account, the proposed force-based traffic simulation framework updates vehicle's motions in both the Frenet coordinates and the Cartesian coordinates. With the way-points from users, lane-level navigation can be generated by reference path planning. With a given keyframe, the coarse-to-fine optimization is proposed to efficiently generate the plausible trajectory which can satisfy the spatio-temporal constraints. At first, a directed state-time graph constructed along the reference path is used to search for a coarse-grained trajectory by mapping the keyframe as the goal. Then, using the information extracted from the coarse trajectory as initialization, adjoint-based optimization is applied to generate a finer trajectory with smooth motions based on our force-based simulation. We validate our method with extensive experiments.


\end{abstract}

%\begin{figure*}[ht]
%    \centering
%    \includegraphics[width=0.95\textwidth]{figures/case_straight.pdf}
%    \caption{An example of controlled simulation with a keyframe set on the adjacent lane. }
%    \label{fig:userstudy}
%\vspace{-0.45cm}
%\end{figure*}

\section{Introduction}
\label{section:introduction}

Traffic simulation has received increasing attention due to the development of computer games, film industry, urban planning, autonomous vehicle driving \cite{li2019aads}, etc. A reliable traffic simulator that can generate high-fidelity virtual traffic data is thus valuable \cite{chao2020survey}.

While realistic traffic flows can be simulated via a variety of simulators~\cite{krajzewicz2012recent, adnan2016simmobility, fellendorf1994vissim, dosovitskiy2017carla}, editing or imposing specific space-time constraints on vehicles is difficult. Meanwhile, the capability of interactively editing vehicle trajectories is needed when traffic scenes need to be simulated with specific vehicles controlled/showing a pre-defined driving behavior. As a result, currently traffic simulation editing has to be based on labor-intensive manual tuning of simulation parameters and exhaustive trial-and-error runs of simulators. Some attempts have been made to address such a limitation, e.g. by allowing users to manually generate desired trajectories or rare traffic events observed less frequently in previous methods or datasets~\cite{han2022traedits}. However, they do not address the spatio-temporal nature of the editing constraints, e.g. specifying a certain vehicle to arrive a certain position at a pre-defined moment. Recently, traffic reconstruction methods~\cite{van2007kinodynamic, sewall2010virtualized} provide a potential solution via optimization with respect to the space-time constraints. But they deteriorate the trajectory quality, e.g. discontinuous and implausible trajectories, and incur large computational costs which renders them unscalable.

To ensure the pluasibility of edited trajectories, TraEDITS~\cite{han2022traedits} integrates a data-driven traffic simulation module inspired by~\cite{ren2019heter}. Data-driven methods can utilize pre-recorded traffic data to generate realistic behaviors. However, there is a fundamental challenge to combine data-driven methods with optimization-based methods which are widely used in traffic simulation. The core reason is the simulation process of optimization-based methods is intrinsically non-differentiable, making gradient-based learning infeasible. It is worth noting that the social force model, which is widely used in crowd animation, has also shown fantastic potential in traffic simulation recently. A unified force-based framework \cite{chao2019force} and a simplified force-based framework \cite{han2021simplified} are successively proposed to simulate mixed traffic scenarios with different types of agents. Because agent's dynamics are explicitly expressed with derivable formulas, optimization based on force models is more practical. However, current force-based traffic simulation methods can only provide scenarios with simple straight lanes, and vehicle's motions are restricted strongly to the lane shapes.

To generate traffic trajectories based on the force-based models with given spatio-temporal constraints, keyframing is a practical and effective technique for controlling coarse results in physically-based simulation. The adjoint method, as one of the popular methods in optimal control for gradient computation, has been introduced in fluid simulation \cite{mcnamara2004fluid} and general particle dynamics \cite{wojtan2006keyframe}, to satisfy given keyframe constraints. However, the results of the adjoint optimization are highly dependent on the initialization. Bad initialization can slow down or even prevent gradient descent from achieving convergence. %However, it is difficult to directly apply adjoint method for traffic simulation controlling, since the complex environments and interactions in traffic scenarios have a great effect on gradient-based optimization especially when the initialization is inappropriate. 

%To address the above challenges, 
%None of the above mentioned methods is able to fulfil all of our requirements perfectly. Firstly, an interactive trajectory editing method which allows users to edit vehicles involving both spatial and temporal constraints is necessary to make editing process more controllable and convenient, while existing frameworks are able to provide spatial edits only. Secondly, a force-based traffic simulation method which is robust enough to receive user's arbitrary edits and generate desired and realistic results is needed, while current methods can only simulate vehicles in scenarios with simple straight lanes. Thirdly, to combine force-based model with spatial-temporal controls, an adjoint-based optimization with gradient computation needs to further adapt to the complex environments and interactions in traffic simulation and guarantee algorithm performance.

None of the prior methods can fulfill all of our requirements perfectly due to the following challenges. Firstly, the current force-based traffic simulation frameworks constrain vehicles to drive along straight lanes and can hardly meet the demands of arbitrary settings or edits given by users. Secondly, there is no proper method for optimizing traffic trajectories with spatio-temporal keyframe control. State-time space search becomes extremely time-consuming if we need smooth traffic behaviours, while gradient-based optimization like the adjoint method may decelerate convergence or generate implausible trajectories due to poor initialization. The optimization process needs to be adapted to efficiently generate plausible traffic behaviours. 

%Secondly, the performance of gradient-based optimization like adjoint method strongly depends on the initialization of the parameters to be optimized. Bad initialization may decelerate convergence of gradient descent, and even generate unacceptable results. The optimization process needs to adapt to the complex environments and interactions in traffic simulation. 

To address the above challenges, we provide a novel traffic trajectory editing method that allows users to control vehicle's motions with spatio-temporal keyframes. Vehicles are updated by force-based models using path coordinates to make them more maneuverable rather than being strongly restricted to straight lanes. A coarse-to-fine optimization process is presented to find optimal keyframe controls of the simulation appropriately. The main contributions of this work are as follows:
\begin{itemize}
\item A traffic trajectory editing approach, which allows users to specify keyframes to regulate vehicles and generate desired trajectories that can meet the spatio-temporal constraints while the traffic behaviours can remain plausible.
\item A novel force-based framework that decouples vehicle's motions from static lanes and can generate microscopic traffic simulation in complex scenarios including diverse interactions. 
\item A coarse-to-fine optimization process is developed by combining the adjoint method with state-time space search to perform gradient descent effectively and stably to generate smooth trajectories with keyframe controls. 
\end{itemize}

\section{Related Work}
\label{section:relatedwork}

\begin{figure*}[ht]
\setlength{\belowcaptionskip}{-0.2cm}
    \centering
    \includegraphics[width=0.99\textwidth]{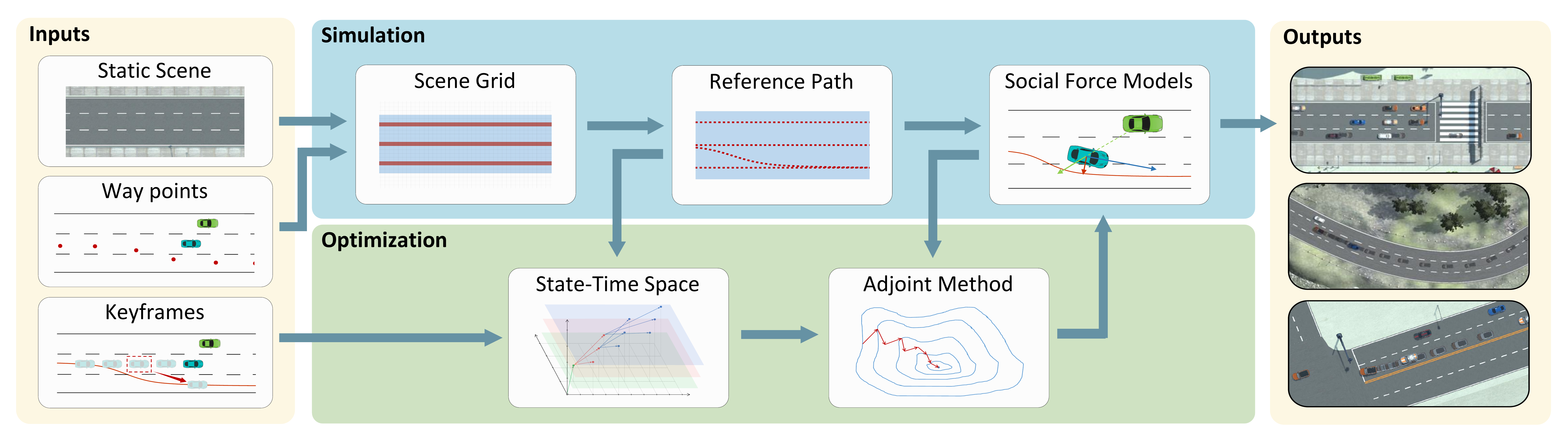}
        \caption{Overview of our spatio-temporal keyframe control of traffic simulation using a coarse-to-fine optimization process. By discretizing static scenarios, we can generate reference paths for vehicles using the given way-points. We simulate the vehicles using our proposed social force models. With given spatio-temporal keyframes, a coarse-grained state-time search and fine-grained adjoint-based optimization are combined to efficiently generate smooth trajectories that can satisfy the keyframe constraints.}
    \label{fig:overview}
%\vspace{-0.45cm}
\end{figure*}

\subsection{Interactive Editing Technology}
Traffic simulation software packages like SUMO \cite{krajzewicz2012recent}, SimMobility \cite{adnan2016simmobility}, Vissim \cite{fellendorf1994vissim} and Carla \cite{dosovitskiy2017carla} can generate traffic flows effectively. However, if users want to edit the results or generate some cases with specific behaviours, they have to tune parameters and run the simulation over and over based on the previous result until it meets the expectation. %, which is quite a frustrating trial-and-error workflow.

The interactive editing concept was proposed to solve the similar problem in crowd animation at the very beginning. Cage-based deformation was introduced to edit large-scale crowd animation interactively \cite{kim2014interactive}. Similar results can be achieved based on mesh deformation \cite{zhang2020crowd}. Users can also control the simulation in real-time by drawing sketches as reference paths or obstacles \cite{montana2017sketching}. In traffic simulation, a human-in-the-loop framework is proposed, which allows users to generate irregular and diverse traffic trajectories \cite{han2022traedits} by generating self-defined reference paths or modifying vehicle's attributes. However, this method only allows users to edit in space dimension, and the temporal constraints like specific arriving time are not supported explicitly. Our method allows users to assign spatio-temporal keyframes to regulate vehicle's motions and generate desired trajectories.

\subsection{Traffic Simulation}
%Traffic simulation has received increasing attention due to the development of computer games, film industry, urban planning, autonomous vehicle driving, etc. 

In computer graphics, data-driven methods are used to generate realistic traffic flows based on pre-captured traffic data from video, LiDAR, GPS, and other available sensors. Spatio-temporal data from in-road sensors is provided to reconstruct traffic flows \cite{wilkie2013flow, li2017city}. Based on texture synthesis and cage-based deformation, appropriate traffic flows can be populated in any road network with given samples \cite{chao2017realistic}. A data-driven optimization-based method is proposed to simulate heterogeneous multi-agent systems \cite{ren2019heter}. However, it is difficult to combine data-driven methods with numerical optimization if we want to add extra spatio-temporal constraints through the generated trajectories since it is difficult to compute the derivative of data-driven simulation processes.

On the other hand, the social force model, which is widely used in crowd animation \cite{helbing1995social, helbing2000simulating, alahi2016social, cosgun2016anticipatory}, has recently shown great potential in traffic simulation due to its high computational effectiveness and flexibility. In order to consider all possible agents in a realistic urban environment, a unified force-based framework is proposed which can describe various interactions among different types of agents \cite{chao2019force}. Based on an object-oriented concept, a simplified model is proposed to parameterize adjustable coefficients for different agents, make parameter tuning more intuitive and improve the scalability of the framework \cite{han2021simplified}. With real-world traffic datasets, the coefficients in the force model can also be calibrated automatically with adaptive genetic algorithm \cite{chao2021calibrated}. However, these methods only provide simulation scenarios with straight lanes, and vehicles' behaviours are strongly restricted and difficult to extend to complex scenarios. We develop a more viable force-based framework to simulate vehicles in different environments and generate diverse behaviours.

\subsection{Keyframe Control Animation}

Keyframe is a classical technique for giving fine-grained controls manually to coarse results in physically-based simulation such as fluid, smoke and cloth. As one of the popular methods of optimal control problems involving gradient computation, the adjoint method is introduced to control fluid simulation \cite{mcnamara2004fluid}, general particle systems \cite{wojtan2006keyframe} as well as elastic motion editing \cite{li2013interactive}. However, the gradient-based methods greatly rely on the initialization, and are improper to be applied to optimize traffic simulation directly. Specifically, traffic reconstruction methods can respect keyframe constraints via searching for paths by mapping the keyframes as the goals in state-time space \cite{van2007kinodynamic, sewall2010virtualized}, but their performance and generated results strongly depend on the space discretization timestep. Therefore, we combine the adjoint method with state-time space search and propose a coarse-to-fine optimization process that can generate plausible traffic simulation results effectively with given keyframes.

%Keyframe is a classical technique for giving fine-grained controls manually to coarse results in physically-based simulation such as fluid, smoke and cloth. As one of the popular methods of optimal control problem involving gradient computation, adjoint method is introduced to control fluid simulation \cite{mcnamara2004fluid}, general particle systems \cite{wojtan2006keyframe} as well as elastic motion editing \cite{li2013interactive}. However, gradient-based optimization greatly relies on the initial values of the parameters to be optimized. Bad initialization may decelerate convergence of gradient descent. The complex environments and interactions are great challenges to apply adjoint method in traffic simulation. A method proposed for traffic reconstruction can also respect keyframe constraints via spanning state-time space \cite{sewall2010virtualized}, but its performance and generated results strongly depend on the space discretization resolution. Therefore, we combine adjoint method with state-time space search and propose a coarse-to-fine optimization process, which can generate plausible traffic simulation results effectively with given keyframes.

\section{Method}

The overview of our proposed method is demonstrated in Fig. \ref{fig:overview}. We first introduce the Frenet coordinates and our static scenarios representation (Section \ref{section:scenes}).Then we show the force-based traffic simulation including social forces calculation (Section \ref{subsection:forces}) and reference path planning (Section \ref{subsection:refplanning}). In order to provide keyframe controls with spatio-temporal constraints, we come up with a coarse-to-fine optimization process (Section \ref{section:optimization}). The state-time graph is constructed (Section \ref{subsection:sttconstruct}) to search for coarse-grained trajectories as initialization (Section \ref{subsection:sttsearch}, \ref{subsection:adjinitial}), and the adjoint method is further applied to obtain smooth trajectories and plausible behaviours (Section \ref{subsection:adjoptimize}).

%Then we find a trajectory in the coarse-grained state-time space (Section \ref{section:sttspace}). We discretize the state-time space along any paths (Section \ref{subsection:sttdiscretize}) and search a sequence of state-time points respecting the given spatial-temporal keyframes (Section \ref{subsection:sttplanning}). These points will further act as initial values in adjoint-based optimization (Section \ref{section:adjiont}) to obtain a refined smoother trajectory.  

\subsection{Frenet Coordinates and Scenario Representation}
\label{section:scenes}

We perform our framework in both the Cartesian coordinates and the Frenet coordinates. Frenet frame is a more intuitive way to describe the vehicle state in the lane. A Frenet coordinate $[s, d]$ consists of the longitudinal displacement along the lane and the lateral deviation relative to the lane center. We use $\textbf{\emph{p}}=[x, y]$, $\hat{\textbf{\emph{p}}}=[s, d]$, $\textbf{\emph{v}}=[v^x, v^y]$ and $\hat{\textbf{\emph{v}}}=[v^s, v^d]$ to represent positions and velocities in the Cartesian coordinates and the Frenet coordinates, respectively. Similarly, we use notations $\textbf{\emph{f}}=[f^x, f^y]$ and $\hat{\textbf{\emph{f}}}=[f^s, f^d]$ to represent forces or other attributes in these two coordinate systems.

Specifically in our work, \emph{Lane} is the static areas where vehicles can drive, usually defined segment-by-segment with specific lane width in the configuration files. \emph{Path} is a sequence of points linking a starting point to a destination, and we interpolate them by cubic spline. Namely, the center line of each \emph{Lane} with a specific driving direction is a \emph{Path}. We use $\mathcal{L}$ and $\mathcal{P}$ to represent a \emph{Lane} and a \emph{Path}, respectively.

%\begin{figure}[t]
%    \centering
%    \includegraphics[width=0.9\columnwidth]{figures/lane approximation.pdf}
%    \caption{Lane approximation using capsule in the grid map. The white areas are undrivable, the blue areas are drivable and the red areas are center lines of the lanes. }
%    \label{fig:laneapproximate}
%\end{figure}

Given a certain scene configuration, the lanes in the scene are always determined. So we initialize the reference path set as:
\begin{equation}
\label{eq:initpath}
\begin{aligned}
    &\mathcal{P}^* = {\mathcal{P}^*_{topo}}\cup{\mathcal{P}^*_{user}}, \\
    &\mathcal{P}^*_{topo} = DFS\left(\mathcal{L}^*\right), \\
    &\mathcal{P}^*_{user} = \varnothing,
\end{aligned}
\end{equation}
where $\mathcal{P}^*$ is the reference path set containing all the available reference paths in the scene, and $\mathcal{L}^*$ is the lane set containing all the lanes defined in the configuration file. $\mathcal{P}^*_{topo}$ and $\mathcal{P}^*_{user}$ represent the reference paths determined by lanes' topology and user settings, respectively. Each path in $\mathcal{P}^*_{topo}$ is a unique link from the origin of a lane without in-comings to the end of a lane without out-goings. $DFS$ represents the depth-first search function. 

In accordance with \cite{han2022traedits}, we generate a grid map for the scene in 2D space with a given resolution. The lanes are embedded into the grid nodes by capsule-like shape approximation for every segment, and we label the grid cells of undrivable areas, drivable areas and lane centers with different values. The 2D grid map will be used to plan the reference paths with way-points from users and store vehicles' information to accelerate real-time neighbor search during the simulation.

\subsection{Force-based Traffic Simulation}
\label{section:fbsimulation}

\subsubsection{Force Calculation}
\label{subsection:forces}

We model three factors which heavily influence a vehicle's state: self-motivation, surrounding neighbors and the environment. The state of a vehicle at time $t$ is denoted as $[\hat{\textbf{\emph{v}}}_t, \hat{\textbf{\emph{p}}}_t, \textbf{\emph{v}}_t, \textbf{\emph{p}}_t, \theta_t, \hat{\textbf{\emph{v}}}_{o,t}, \mathcal{P}_{k}]$. $\hat{\textbf{\emph{v}}}_t, \textbf{\emph{v}}_t\in\mathbb{R}^2$ represents its velocity in Frenet coordinates and Cartesian coordinates, respectively. $\hat{\textbf{\emph{p}}}_t$,  $\textbf{\emph{p}}_t\in\mathbb{R}^2$ represent its position in Frenet coordinates and Cartesian coordinates, respectively. $\theta_t\in\mathbb{R}$ represents its orientation by Euler angles. $\hat{\textbf{\emph{v}}}_{o,t}\in\mathbb{R}^2$ represents the free-flow desired velocity. $\mathcal{P}_{k}\in\mathcal{P}^*$ represents the reference path $k$ it follows. The vehicle's attributes in the Frenet coordinates and the Cartesian coordinates can be interconverted by the cubic spline function $S_k$ of its reference path. The dynamics of a vehicle are formulated as:
\begin{equation}
\label{eq:dynamics}
\begin{aligned}
    %& \hat{\textbf{\emph{f}}}_{t} = \hat{\textbf{\emph{f}}}_{o,t} + \hat{\textbf{\emph{f}}}_{k,t} + \sum_{j\in \mathcal{N}_{t}}\hat{\textbf{\emph{f}}}_{j,t}, \\
    & \hat{\textbf{\emph{f}}}_{t} = \hat{\textbf{\emph{f}}}_{o,t} + \hat{\textbf{\emph{f}}}_{k,t} + S_k \left( \sum_{j\in \mathcal{N}_{t}}{\textbf{\emph{f}}}_{j,t} \right), \\
    & \hat{\textbf{\emph{v}}}_{t+1} = \hat{\textbf{\emph{v}}}_t + \frac{\hat{\textbf{\emph{f}}}_{t}}{m} \cdot \Delta{t}, \\
    & \hat{\textbf{\emph{p}}}_{t+1} = \hat{\textbf{\emph{p}}}_t + \hat{\textbf{\emph{v}}}_{t+1} \cdot \Delta{t}, \\
    [ & \textbf{\emph{v}}_{t+1}, \textbf{\emph{p}}_{t+1}, \theta_{t+1}] = S_k\left( \hat{\textbf{\emph{p}}}_{t+1} \right), 
\end{aligned}
\end{equation}
where $\hat{\textbf{\emph{f}}}_{t}$ represents the total force on the vehicle in Frenet coordinates, which is a combination of the self-motivated force $\hat{\textbf{\emph{f}}}_{o,t}$, the path keeping force $\hat{\textbf{\emph{f}}}_{k,t}$ from its reference path $\mathcal{P}_k$ and the collision avoidance forces ${\textbf{\emph{f}}}_{j,t}$ exerted by the neighbor vehicle $j$ in its neighbors set $\mathcal{N}_{t}$. Specifically, the collision avoidance forces are computed in Cartesian coordinates at first. $m$ represents the vehicle's mass. The velocity and the position in Frenet coordinates are updated by the forces. The velocity and the position in Cartesian coordinates as well as the orientation are obtained using the cubic spline function $S_k$. $\Delta{t}$ is the timestep used in the simulation.

\textbf{Self-motivated force:} We assume that each vehicle has a desired velocity to travel at when it is not constrained by the presence of neighbor vehicles. Different from the previous methods \cite{chao2019force, han2021simplified, chao2021calibrated}, we formulate the self-motivated force as:
\begin{equation}
\label{eq:selfmotivatedforce}
    \hat{\textbf{\emph{f}}}_{o,t} = \omega_o m \left( \frac{2}{1+e^{\hat{\textbf{\emph{v}}}_{o,t} - \hat{\textbf{\emph{v}}}_{t}}} - 1 \right) \hat{\textbf{\emph{a}}},
\end{equation}
where $\omega_o$ is a corresponding weight, and $\hat{\textbf{\emph{a}}}$ is the maximum acceleration of the individual in Frenet coordinates. Such formulation can make vehicles gradually change and keep their velocity to the desired $\hat{\textbf{\emph{v}}}_{o,t}$, while the derivative of the expression exists at each point in its domain. % It is important because the adjoint-based optimization introduced in the later section depends on the differentiability.

\textbf{Path keeping force:} Typically drivers tend to drive along the lane center due to safety and traffic rules in real world traffic. As mentioned, we consider the center line of a lane with a driving direction as a path, so the lane keeping behaviors can also be regarded as path keeping. We define the path keeping force as an attractive force from the path:
\begin{equation}
\label{eq:pathkeepingforce}
    %\hat{\textbf{\emph{f}}}_{k} = \omega_{k} \left| d \right|,
    \hat{\textbf{\emph{f}}}_{k} = 
    \left\{
        \begin{array}{lr}
        \omega_{k} \left| d_t \right| \textbf{\emph{u}}_k, & \left| d_t \right| \geq \frac{1}{2}(w_l - w_v) \\
        \textbf{\emph{0}}, & otherwise
        \end{array}, 
    \right.
\end{equation}
where $\omega_k$ is a corresponding weight, $d_t\in\hat{\textbf{\emph{p}}}_t$ is the current lateral displacement of the vehicle related to the path, and $\textbf{\emph{u}}_k$ is the unit vector pointing from the vehicle to the path. $w_l$ is the lane width, and $w_v$ is the vehicle's width. As shown, path keeping force is active only if a vehicle's deviation from its reference path exceeds the threshold in order to prevent the vehicle oscillating around the path.

\textbf{Collision avoidance force:} A vehicle should avoid colliding with others who are too close to it. We formulate such behaviour as a point-to-point repulsive force that effects between the vehicle and its surrounding neighbors. The collision avoidance force between a vehicle and its neighbor $j$ is defined as:

\begin{equation}
\label{eq:collisionavoidforce}
\begin{aligned}
    & {\textbf{\emph{f}}}_{j} = \omega_c \frac{a  \cdot b }{\left(1+c||\textbf{\emph{p}}_{j,t} - \textbf{\emph{p}}_t||  \right)^2}\textbf{\emph{u}}_c, \\
    & a = \left\{
        \begin{array}{lr}
        cos\phi, & \phi \leq \frac{\pi}{4} \\
        0, & otherwise
        \end{array}, 
    \right. \\
    & b = s_0 + ||\textbf{\emph{v}}_t || T_0 + \frac{||\textbf{\emph{v}}_t||\cdot || \textbf{\emph{v}}_{j,t} - \textbf{\emph{v}}_t ||}{2||\hat{\textbf{\emph{a}}}||}, \\
    & c = \frac{1}{s_0},
\end{aligned}
\end{equation}
where $\omega_c$ is a corresponding weight,  $a$ is a visual factor, $b$ and $c$ are parameterized coefficients, and $\textbf{\emph{u}}_c$ is the unit vector pointing from $j$ to the vehicle. We formulate the coefficients inspired by the intelligent driver model (IDM) \cite{treiber2000congested}. $\phi$ is the angle between the vehicle's moving direction and the direction of the vehicle pointing to $j$. $\textbf{\emph{v}}_{j,t}$ and $\textbf{\emph{p}}_{j,t}$ is the velocity and the position of $j$ in Cartesian coordinates. $s_0$ and $T_0$ are the safety space headway between two vehicles and the reacting time for the vehicle to brake, which are both constant for a certain individual. The neighbor set $\mathcal{N}_t$ can be updated at each frame effectively by the 2D grid map, where the search range is 100 × 100 grid nodes in our implementation.

\begin{figure}[t]
\setlength{\belowcaptionskip}{-0.2cm}
    \centering
    \includegraphics[width=0.9\columnwidth]{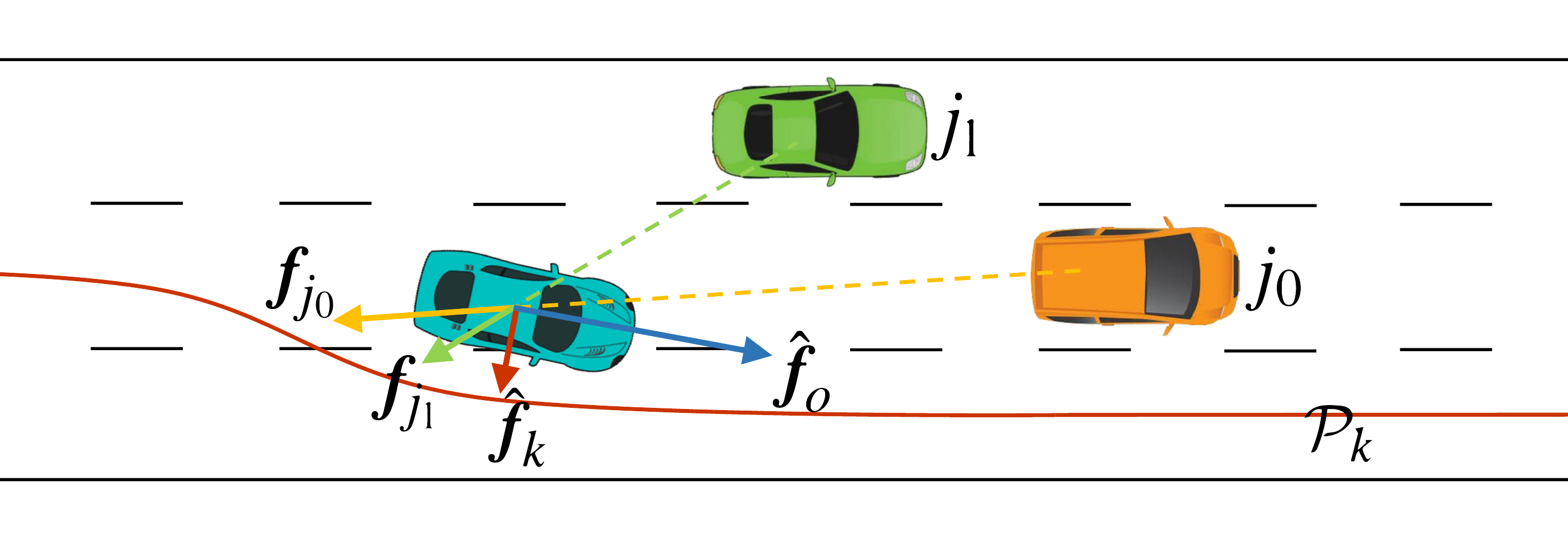}
    \caption{An example of the forces on a vehicle. The vehicle follows the reference path $\mathcal{P}_k$ and has two neighbors, where $j_0$ is in the same lane and $j_1$ is in the adjacent lane. }
    \label{fig:forces}
\end{figure}

Based on the force models, our traffic simulation can generate smooth traffic behaviours such as acceleration/deceleration, lane keeping and lane changing. Since we update the vehicles along their reference paths, our framework can simulate the traffic in complex scenarios. We give an example of the forces on a driving vehicle in Fig. \ref{fig:forces}. %It also becomes possible to optimize through the whole trajectory in real-time thanks to its high computational effectiveness. We give an example of forces on a driving vehicle in Fig. \ref{fig:forces}. 
% and allows users to edit each individual to generate desired trajectories

\subsubsection{Reference Path Planning}
\label{subsection:refplanning}
In addition to the paths extracted from lane centers, new reference paths can be created by clicking a sequence of key points in the scene. The given key points are mapped to the specific nodes on the constructed 2D grid map, and then the whole path is planned segment-by-segment, where the first node of each segment is considered as the start and the second one is the terminal. We use the A* algorithm to plan the path and define the heuristic function as:
\begin{equation}
\label{eq:heuristic}
    h(\textbf{\emph{n}}) = || \textbf{\emph{n}} - \textbf{\emph{n}}_{goal} || + \mu_a  \cdot e^{\left( \mu_b \cdot \emph{sign} \right)},
\end{equation}
where $|| \textbf{\emph{n}} - \textbf{\emph{n}}_{goal} ||$ is the Euler distance between the current node and the terminal, and $\mu_a$ and $\mu_b$ are adjustable coefficients. $sign\in[0,1,2]$ is the sign filled in different types of nodes we metioned in Section \ref{section:scenes}, where $0$ represents unreachable area, $1$ represents drivable area and $2$ represents lane center. The second term on the right-hand side of the equation can make the planning tend to search along with lane centers.

We post-process the planning results by down-sampling, Gaussian smoothing and interpolating with cubic spline. Finally the user-defined path $\mathcal{P}$ becomes available for vehicles to follow, denoted as $\mathcal{P}^*_{user} = \mathcal{P}^*_{user} \cup [\mathcal{P}]$. Specifically, once a vehicle's reference path is changed, its state values represented in Frenet coordinates need to be updated according to the new cubic spline function.

%Specifically, the reference path of a vehicle can be changed during the simulation because of user's manual settings or lane changing decisions. Once a vehicle's reference path is changed, its state values represented in Frenet coordinates have to update according to the new cubic spline function. % In our simulation, lane changing only occurs when a vehicle is blocked by the front leader and there are no neighbors in the adjacent lanes. Once the vehicle's reference path is changed, its state values that represented in Frenet coordinates have to update according to the new cubic spline function.

%\subsection{Coarse Search in State-Time Space}
%\label{section:sttspace}
%So far, users can control the vehicle 

%\subsubsection{State-Time Space Discretization}
%\label{subsection:sttdiscretize}

%\subsubsection{Coarse Trajectory Search}
%\label{subsection:sttplanning}

%\subsection{Trajectory Refinement with Adjoint Method}
%\label{section:adjiont}

\subsection{Spatio-temporal Keyframe Control}
\label{section:optimization}
%So far, users can control vehicles' trajectories using self-defined reference paths. 
In order to provide further keyframe controls with both spatial and temporal constraints, the adjoint method is applied since it is proven to be effective in optimal control. We can find a set of optimal controlling desired speeds and corresponding controlling forces, and make vehicle's behaviours satisfy the keyframe constraints based on our force-based simulation.

%However, gradient-based optimization greatly relies on the initial values of the parameters to be optimized. Bad initialization may decelerate convergence of gradient descent, or even lead to unacceptable results. Therefore, we perform another optimization in a coarse-grained state-time space to find a coarse trajectory at first, and then regard it as initialization to further optimize and improve the quality of the behaviours. The pseudo-code of our coarse-to-fine optimization is described in Algorithm \ref{alg:optimization}, and we will introduce the details in the following sections.

However, gradient-based optimization greatly depends on the initial values of the parameters to be optimized. Bad initialization may decelerate or even prevent gradient descent from achieving convergence. Therefore, we perform another optimization in a discrete state-time space to find a coarse-grained trajectory at first, and then treat it as initialization to further optimize and improve the behaviours. The pseudo-code of our coarse-to-fine optimization is described in Algorithm \ref{alg:optimization}, and we will introduce the details in the following sections.

\begin{algorithm}[t]
\setlength{\belowdisplayskip}{-0.25cm}
\caption{Coarse-to-fine optimization}
\label{alg:optimization}
\begin{algorithmic}[1]

\REQUIRE ~~\\ %算法的输入参数：Input
    State-time graph for a vehicle along its reference path; \\
    %The start keyframe $[s_{start}, v^s_{start}, t_{start}]$; \\
    %The goal keyframe $[s_{goal}, v^s_{goal}, t_{goal}]$; \\
    $K$ setting keyframes $Q= [\Tilde{\textbf{\emph{q}}}_0, \Tilde{\textbf{\emph{q}}}_1, ..., \Tilde{\textbf{\emph{q}}}_K]$; \\
    The maximum iteration number $N$ for the adjoint method; \\
    Learning rate $\alpha$, exponential decay rates $\beta_0$, $\beta_1$; \\
\ENSURE ~~\\ %算法的输出：Output
    Trajectory $\mathcal{T}$ under keyframes control; \\~\\

\STATE Initialize controlling desired speeds $V \Leftarrow \varnothing$, controlling forces $F \Leftarrow \varnothing$; 
\FOR{$i=0 $ to $K-1$}
    \STATE Start state-time node $[s_{start}, v^s_{start}, t_{start}] \Leftarrow \Tilde{\textbf{\emph{q}}}_i$;
    \STATE Goal state-time node $[s_{goal}, v^s_{goal}, t_{goal}] \Leftarrow \Tilde{\textbf{\emph{q}}}_{i+1}$;
    \STATE Find a coarse trajectory in state-time graph with A* algorithm, $\mathcal{T}_c \Leftarrow \mathcal{A}^*([s_{start}, v^s_{start}, t_{start}], [s_{goal}, v^s_{goal}, t_{goal}])$;
    \STATE Append controlling desired speeds, $V \Leftarrow V\cup[v^s_{o,0}, v^s_{o,1},...]$ extracted from current $\mathcal{T}_c$;
\ENDFOR

\STATE Pad controlling  desired speeds $V$;

\FOR{$i=0$ to $N-1$}
    \STATE Compute corresponding forces with controlling desired speeds, $F \Leftarrow V$;
    \STATE Simulate the current whole trajectory $\mathcal{T}_o$ with $F$ using our force-based traffic simulation algorithm;
    
    \IF{loss decreases}
        \STATE Update $\mathcal{T} \Leftarrow \mathcal{T}_o$;
    \ENDIF
    
    %\STATE Compute gradient of desired speeds using adjoint method, $\nabla{V} \Leftarrow \mathcal{A}djoint(T_o, [\Tilde{\textbf{\emph{n}}}_0, \Tilde{\textbf{\emph{n}}}_1, ..., \Tilde{\textbf{\emph{n}}}_k])$;
    \STATE Compute gradient of desired speeds using the adjoint method, $\nabla{V} \Leftarrow \mathcal{A}djoint(\mathcal{T}_o, V)$;
    
    \STATE Gradient descent, $V \Leftarrow Adam(\nabla{V}, \alpha,\beta_0,\beta_1)$;
    %\STATE Gradient descent by Adam optimizer, $V \Leftarrow Adam(\nabla{V})$;
\ENDFOR

%\STATE Find a coarse trajectory \\ $T_c = Heuristic Search([s_{start}, v^s_{start}, t_{start}], [s_{goal}, v^s_{goal}, t_{goal}])$;
\RETURN Fine trajectory $\mathcal{T}$;

\end{algorithmic}
\end{algorithm}

\subsection{Coarse Search in State-Time Space}
\label{section:sttspace}

\subsubsection{State-Time Graph Construction}
\label{subsection:sttconstruct}
We define the state space of a vehicle as a subset of its entire state representation shown in Section \ref{subsection:forces} for simplicity in the following. The state space along a vehicle's reference path in the longitudinal direction is denoted as $[s, v^s]$, where $s\in\hat{\textbf{\emph{p}}}$ and $v^s\in\hat{\textbf{\emph{v}}}$ are the longitudinal displacement and the longitudinal speed, respectively. 

The state space can be discretized into grid nodes by giving another timestep $\Delta \Tilde{t}$, which is much larger than the one we use in traffic simulation. Assume that the acceleration of a vehicle can only be chosen from a discrete set $[-a^s, 0, a^s]$ where $a^s\in\hat{\textbf{\emph{a}}}$ is the maximum longitudinal acceleration. According to the state dynamics shown in Eq. \ref{eq:dynamics}, the intervals along the $v^s$-axis and the $s$-axis are $\Delta{v^s}=a^s\Delta{\Tilde{t}}$ and $\Delta{s}=\frac{1}{2}a^s(\Delta{\Tilde{t}})^2$. Therefore, starting from a given state node $[s, v^s]$, there should be three reachable state nodes 
after $\Delta{\Tilde{t}}$: $[s+(2\frac{v^s}{\Delta{v}^s}+1)\Delta{s}, v^s + \Delta{v}^s]$ for acceleration, $[s+2\frac{v^s}{\Delta{v}^s}\Delta{s}, v^s]$ for speed maintenance and $[s+(2\frac{v^s}{\Delta{v}^s}-1)\Delta{s}, v^s - \Delta{v}^s]$ for deceleration. As a result, the behaviours along the reference path can be represented as a directed graph with finite combinations of reachable state nodes. 

\begin{figure}[t]
\setlength{\belowcaptionskip}{-0.2cm}
    \centering
    \includegraphics[width=0.9\columnwidth]{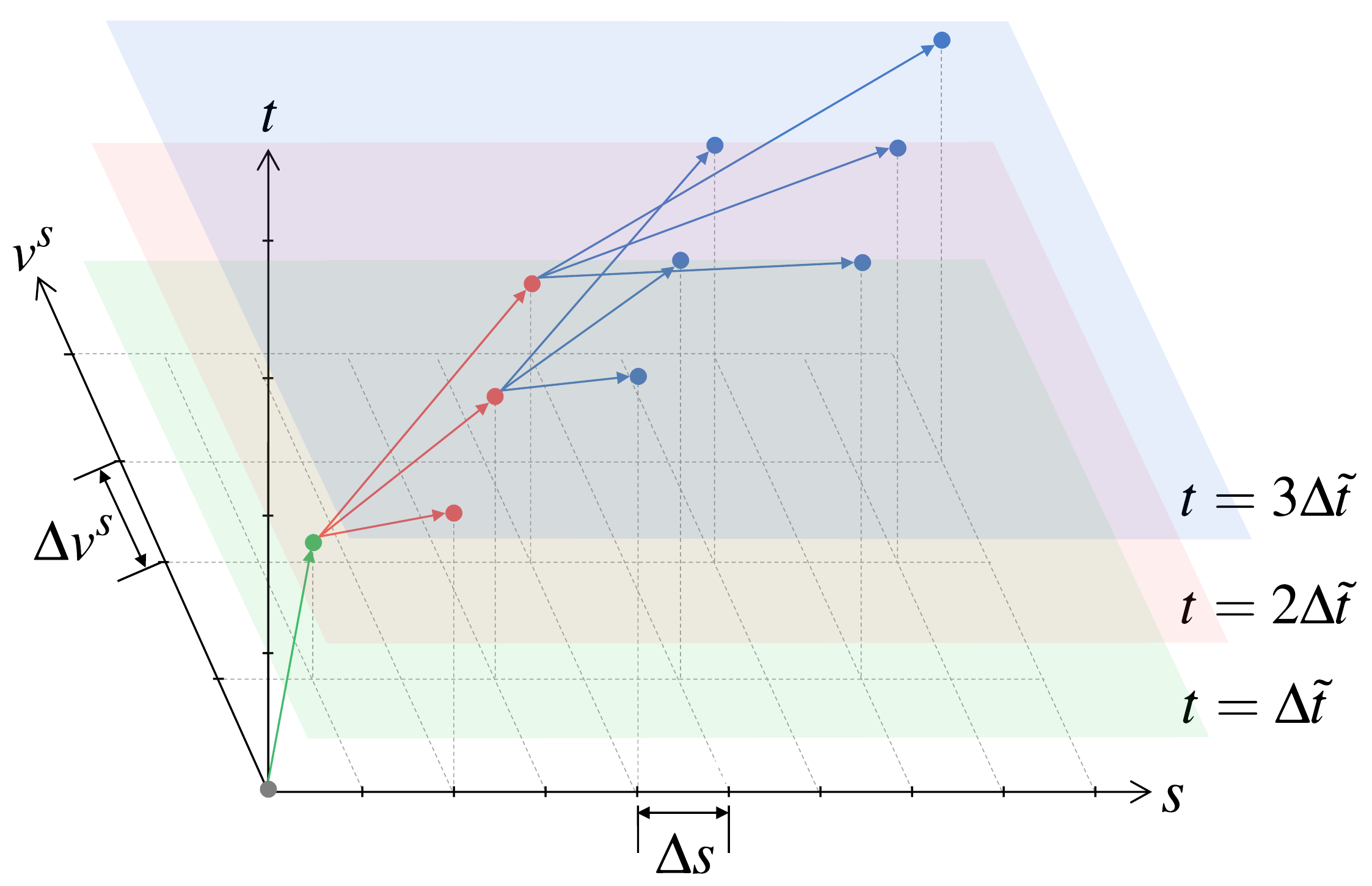}
     \caption{A part of discretized state-time space and directed state-time graph. The reachable nodes are marked by the dots with different colors laying on the corresponding $s$-$v^s$ planes, representing time at $\Delta\Tilde{t}$ (green), $2\Delta\Tilde{t}$ (red) and $3\Delta\Tilde{t}$ (blue). The transitions from parent nodes denoted by arrows are also colorized. }
    \label{fig:sttspace}
\end{figure}

The state-time space of a vehicle is the state space augmented by the time dimension. Similarly, we can also construct a directed graph in the state-time space that contains transitions from given state-time nodes $\Tilde{\textbf{\emph{q}}}=[s, v^s, t]$ to their corresponding three reachable state-time nodes. Therefore, the given keyframes can be mapped to specific state-time nodes, and we need to find a trajectory from the start node $[s_{start}, v^s_{start}, t_{start}]$ to the goal $[s_{goal}, v^s_{goal}, t_{goal}]$ through the spanning graph. A part of discretized state-time space and state-time graph are demonstrated in Fig. \ref{fig:sttspace}.

The state-time space and the corresponding directed state-time graph for any vehicle are identical with the same discretization process. In practice, we only need to explicitly construct one directed state-time graph and reuse it for different reference paths.

\subsubsection{Coarse Trajectory Search}
\label{subsection:sttsearch}
Finding a trajectory through the state-time graph with given a start and goal can be seen as finding an optimal path in the special 3D solution space. We also use the A* algorithm to search for it. The heuristic function for a certain node is defined as:
\begin{equation}
\label{eq:sttheuristic}
\begin{aligned}
    & \Tilde{h}(\Tilde{\textbf{\emph{q}}}) = \omega_d \Tilde{h}_d(\Tilde{\textbf{\emph{q}}}) + \omega_a \Tilde{h}_a(\Tilde{\textbf{\emph{q}}}), \\
    & \Tilde{h}_d(\Tilde{\textbf{\emph{q}}}) = \sqrt{ | s - s_{goal} |^2 + | v^s - v^s_{goal} |^2 + | t - t_{goal} |^2 },\\
    & \Tilde{h}_a(\Tilde{\textbf{\emph{q}}}) = \frac{|v^s - v^s_{parent}|}{\Delta{\Tilde{t}}} ,\\
\end{aligned}
\end{equation}
where $\Tilde{h}_d$ is the distance between the state-time node and the goal, and $\Tilde{h}_a$ is the variation of speed compared with its parent state, making the vehicle accelerate or decelerate as infrequently as possible. $\omega_d$ and $\omega_a$ are corresponding weights.

Moreover, other moving vehicles can be pre-converted to static obstacles in the state-time space \cite{sewall2010virtualized}. During the A* search, the nodes occupied by other vehicles are marked as unreachable from their parent nodes. The nodes that exceed allowed speed, path length or maximum time duration are also prohibited. However, such constraints may raise failure that we can not find a trajectory that matches the keyframes. To ensure that the A* algorithm always produces an output, we will return the trajectory that can reach the node closest to the goal and has the minimal heuristic value when the search can not reach the true goal.

Obviously, this generated trajectory is implausible because of the large timestep used to construct the state-time graph as well as the discrete acceleration options. Though using a small timestep or adding more alternative acceleration values can generate better results, they will also make space discretization and state-time search become extremely time-consuming because of the exponential rise of the number of state-time nodes. Therefore, we will use the adjoint method to refine the behaviours along the coarse trajectory depending on our force-based simulation.

\subsection{Trajectory Refinement with Adjoint Method}
\label{subsection:adjoint}

\subsubsection{Initialization Using Coarse Trajectory}

\label{subsection:adjinitial}
%We use $\mathcal{T}$ to represent a trajectory. After searching a coarse trajectory $\mathcal{T}_c$ in state-time space, we use the speeds $V=[v_{0}^s, v_{1}^s, ...]$ extracted from $\mathcal{T}_c$ as the initialized controlling desired speeds, and further find a set of optimal controlling desired speeds to optimize vehicle's behaviours based on our self-motivated force model in Eq. \ref{eq:selfmotivatedforce}.

We use the notation $\mathcal{T}$ to represent a trajectory in the following. After searching a coarse trajectory $\mathcal{T}_c$ in state-time space, we use the speeds extracted from $\mathcal{T}_c$ as the initialized controlling desired speeds $V=[v_{o,0}^s, v_{o,1}^s, ...]$, and further optimize them based on our force models to make the simulation satisfy the given keyframes.

As we mentioned in the last section, different timesteps in traffic simulation and state-time space discretization are used. It will cause different numbers of frames when a certain trajectory is registered by simulation and state-time space. We pad the sequence of controlling desired speeds $V$ extracted from coarse trajectory by linear interpolation to align the number of frames with the trajectory generated by simulation. Then we can use the padded $V$ to calculate the corresponding controlling forces $F$ and trajectory $\mathcal{T}_o$. $\mathcal{T}_c$ and $\mathcal{T}_o$ can be seen as the different representations for a certain trajectory in state-time space and traffic simulation, respectively. Obviously, we have a relationship that $\#\mathcal{T}_o \cdot \Delta{t} = \#\mathcal{T}_c \cdot \Delta{\Tilde{t}}$, where $\#\mathcal{T}_o$ and $\#\mathcal{T}_c$ are the numbers of frames, and both sides are equal to the total driving time duration of this trajectory.

%, and have the relationships
%\begin{equation}
%\label{eq:relationship}
%\mathcal{T}_o \cdot \Delta{t} = \#\mathcal{T}_c \cdot \Delta{\Tilde{t}},
%\end{equation}
%where $\#\mathcal{T}_o$ and $\#\mathcal{T}_c$ are the trajectory lengths of each, and both sides of the equation are equal to the driving time duration of the vehicle.

\begin{table*}[htbp]
\setlength{\belowcaptionskip}{-0.2cm}
\centering
\normalsize
\renewcommand\arraystretch{1.2}
\begin{tabular}{|c|c|c|c|}

%\multicolumn{1}{c}{\textbf{Parameter} & \textbf{Value} & \textbf{Unit} & \textbf{Description}}
\hline
\textbf{Parameter}                   & \textbf{Value}         & \textbf{Unit}     & \textbf{Description}                                                       \\ \hline
$\Delta{t}$               & 0.01          & $s$     & the timestep used in the force-based traffic simulation                                \\ \hline
$\hat{\textbf{\emph{a}}}$ & [5.0, 1.0]    & $m/s^2$ & the maximum acceleration of the vehicles in Frenet coordinates                             \\ \hline
$s_0$                     & 4.0 $\pm$ 1.0 & $m$     & the jam space headway between two vehicles for safety                                  \\ \hline
$T_0$                     & 1.0 $\pm$ 0.5 & $s$     & the reacting time for the vehicles to brake                                                \\ \hline
$w_v$, $w_l$              & 1.8, 3.5      & $m$     & the width of lanes and the vehicles, respectively                                          \\ \hline
$\omega_o$, $\omega_k$, $\omega_c$ &
  1.0, 0.5, 3.0 &
  $-$ &
  the weights for the force calculations in Eq. \ref{eq:selfmotivatedforce}, Eq. \ref{eq:pathkeepingforce} and Eq. \ref{eq:collisionavoidforce}, respectively \\ \hline
$\mu_a$, $\mu_b$          & 20.0, $-1.5$    & $-$       & the coefficients for the heuristic-based path planning in Eq. \ref{eq:heuristic} \\ \hline
$\Delta{\Tilde{t}}$       & 0.5           & $s$     & the timestep used in state-time space discretization                                   \\ \hline
$\omega_d$, $\omega_a$    & 1.0, 2.0      & 
$-$       & the weights for the heuristic-based state-time search in Eq. \ref{eq:sttheuristic}  \\ \hline
$\omega_t$, $\omega_v$ & 1.0, 0.1 & $-$ & the weights for the objective function in the adjoint method in Eq. \ref{eq:adjobjective} \\ \hline

$\alpha$, $\beta_0$, $\beta_1$ & 0.01, 0.9, 0.999 & $-$ & the parameters of Adam optimizer used for gradient descent in Algorithm \ref{alg:optimization} \\ \hline
  
\end{tabular}
\caption{The values of the important parameters used in our experiments.}
\label{tab:parameters}
\end{table*}

\subsubsection{Adjoint-based Optimization}
\label{subsection:adjoptimize}
To measure the difference between the trajectory $\mathcal{T}_o$ with $T$ frames generated by the controlling desired speeds $V$ and a set of keyframes $Q$ specified by users, we define the objective function which we want to minimize as:
\begin{equation}
\label{eq:adjobjective}
\begin{aligned}
    & \Phi(\mathcal{T}_o, V) = \frac{1}{2} \sum_{t=0}^T \left(  \omega_t || \mathcal{T}_{o,t} - Q_t ||^2 + \omega_v ||v^s_{o,t}|| \right), \\
    & s.t. \quad \mathcal{T}_{o,t+1} = G\big(\mathcal{T}_{o,t}, v^s_{o,t}\big), \quad t\in[0,1,...T-1],
\end{aligned}
\end{equation}
%where $\omega_t$ is a weight to emphasize the influence of the state at certain keyframes. $\mathcal{T}_{o,t} = [s_t, v^s_t] $ is the state at time $t$ along the trajectory $\mathcal{T}_o$, and $Q_t$ is the keyframe at time $t$ if there exists. A regularization weighted by $\omega_v$ for the controlling desired speeds is also added to prevent overfitting.
where $\omega_t$ is a weight to emphasize the influence of the state at certain keyframes, $\mathcal{T}_{o,t} = [s_t, v^s_t]$ is the state at time $t$ along the trajectory, and $Q_t$ is the keyframe at time $t$ if there exists. In fact, $\mathcal{T}_{o,t}$ should be the state-time node $[s, v^s, t]$ as our definition, but the time dimension is left out here because we strictly align every timestamp when calculating the objective function. A regularization term weighted by $\omega_v$ is also added to prevent overfitting.

%where $\omega_t$ is a weight to emphasize the influence of the state at certain keyframes, $\mathcal{T}_{o,t} = [s_t, v^s_t] $ is the state at time $t$ along the trajectory $\mathcal{T}_o$ and $Q_t$ is the keyframe at time $t$. In fact, most keyframes except for the start and the goal will only have a weight $\omega_t$ of 0. A regularization weighted by $\omega_v$ for the controlling desired speeds is also added to prevent over fitting.  %The term weighted by $\omega_v$ is to make the desired speeds used to minimize the keyframe deviations as small as possible. 

As shown in Eq. \ref{eq:adjobjective}, the optimization should satisfy a series of time stepping constraints advanced via function $G$, an abbreviation of the state dynamics functions we have already shown in Eq. \ref{eq:dynamics}. According to the adjoint method, we introduce a set of Lagrange multipliers and transform the optimization into
\begin{equation}
\label{eq:adjreverse}
\begin{aligned}
    & \nabla{V} = \frac{d\Phi}{dV} = \sum_{t=0}^T \lambda_t \cdot \frac{\partial G}{\partial v^s_{o,t}} + \frac{\partial \Phi}{\partial V} ,\\
    & \lambda_t = \left\{
        \begin{array}{lr}
        \frac{\partial \Phi}{\partial \mathcal{T}_{o,t}},  & t = T \\
       \lambda_{t+1} \cdot \frac{\partial G}{\partial \mathcal{T}_{o,t}} + \frac{\partial \Phi}{\partial \mathcal{T}_{o,t}}, & t < T
        \end{array},
    \right.
\end{aligned}
\end{equation}
where $\lambda_t$ is the Lagrange multiplier for time $t$, which is also called an adjoint state in the adjoint method. These Lagrange multipliers are calculated by iterating backward in time at first, and then we can obtain the gradient of the controlling desired speeds $V$ by substitution. Finally, the controlling desired speeds are updated by gradient descent algorithm to get a new set of controlling forces which tends to decrease the difference between the simulation trajectory and the given keyframes. We use Adam optimizer for gradient descent in our implementation.

For clarity, we further demonstrate how to solve the terms ${\partial G}/{\partial v^s_{o,t}}$ and ${\partial G}/{\partial \mathcal{T}_{o,t}}$ in the above equations. For the speed component $v^s\in[s, v^s]$ along $\mathcal{T}_{o}$, according to Eq. \ref{eq:dynamics}, the transition function $G$ can be written as:
\begin{equation}
\label{eq:adjspeedtrans}
    %v^s_{t+1} = G(v^s_t, s_t, v^s_{o,t}) = v^s_{t} + \frac{f^s_t}{m} \cdot \Delta{t},
    v^s_{t+1} = G(\mathcal{T}_{o,t}, v^s_{o,t}) = v^s_{t} + \frac{f^s_t}{m} \cdot \Delta{t},
\end{equation}
where $f^s_t$ is the longitudinal component of the total force at time $t$. So we can obtain that
\begin{equation}
\label{eq:adjspeed}
\begin{aligned}
    & \frac{\partial G}{\partial v^s_{o,t}} = \frac{\partial f^s_t}{\partial v^s_{o,t}} \frac{\Delta{t}}{m}, \\
%   & \frac{\partial G}{\partial \mathcal{T}_{o,t}} = \left\{
%        \begin{array}{lr}
%        1 + \frac{\partial f^s_t}{\partial v^s_t} \frac{\Delta{t}}{m}, \\
%        \frac{\partial f^s_t}{\partial s_t} \frac{\Delta{t}}{m},
%        \end{array}.
%    \right.
    & \frac{\partial G}{\partial \mathcal{T}_{o,t}} = 
    \left[
        1 + \frac{\partial f^s_t}{\partial v^s_t} \frac{\Delta{t}}{m}, \frac{\partial f^s_t}{\partial s_t} \frac{\Delta{t}}{m}
    \right].
\end{aligned}
\end{equation}
According to the force models demonstrated in Section \ref{subsection:forces}, we can easily calculate these expressions. In our implementation, we only compute the derivative of self-motivated force. Since the collision avoidance force calculation contains a non-differentiable piecewise function, and the path keeping force has no contribution to the vehicle's longitudinal motions. In a similar way, the two terms for the position component $s\in[s, v^s]$ are
\begin{equation}
\label{eq:adjpos}
\begin{aligned}
    & \frac{\partial G}{\partial v^s_{o,t}} = \frac{\partial f^s_t}{\partial v^s_{o,t}} \frac{(\Delta{t})^2}{m}, \\
%   & \frac{\partial G}{\partial \mathcal{T}_{o,t}} = \left\{
%        \begin{array}{lr}
%        \Delta{t}\left( 1 + \frac{f^s_t}{v^s_t} \frac{\Delta{t}}{m} \right), \\
%        1 + \frac{f^s_t}{s_t} \frac{(\Delta{t})^2}{m},
%        \end{array}.
%    \right.
    & \frac{\partial G}{\partial \mathcal{T}_{o,t}} =
    \left[
        \Delta{t}\left( 1 + \frac{\partial f^s_t}{\partial v^s_t} \frac{\Delta{t}}{m} \right), 1 + \frac{\partial f^s_t}{\partial s_t} \frac{(\Delta{t})^2}{m}
    \right].
\end{aligned}
\end{equation}

After repeating optimization for certain iterations, the final refined trajectory $\mathcal{T}$ can meet the given spatio-temporal keyframe constraints and also gets smoother.

%$\hat{\textbf{\emph{f}}}_{j_{0}}$

%$\hat{\textbf{\emph{f}}}_{j_{1}}$

\section{Experimental Results}

\subsection{Experimental Setup}

The following experiments were implemented on a computer with a 3.60GHz Intel(R) Xeon(R) W-2123 CPU with 8-core processors and 32GB memory. Our source code was implemented in C++, compiled as a x64 dynamic link library and imported into Unity3D for visualization. The values of some important parameters we used in our experiments are shown in Table \ref{tab:parameters}, which were pre-defined in configure files and loaded by the program.

There are three scenarios used in our cases which were manually created by SUMO NetEdit and exported as XML files. The first scenario contains a curvy road with three lanes in the same direction. The second scenario contains a straight road with three lanes in the same direction as well as a crosswalk. The third scenario contains an intersection with a four-lane dual carriageway. The discretization resolution of the 2D grid map is 0.5m × 0.5m for each.

%\begin{figure}[t]
%    \centering
%    \includegraphics[width=\columnwidth]{figures/case bend v1.pdf}
%     \caption{ The original trajectories (top) and the keyframe controlled trajectories (bottom) of the curvy road scneario. }
%    \label{fig:casecurvy}
%\end{figure}

\begin{figure}[t]
    \centering
    \includegraphics[width=\columnwidth]{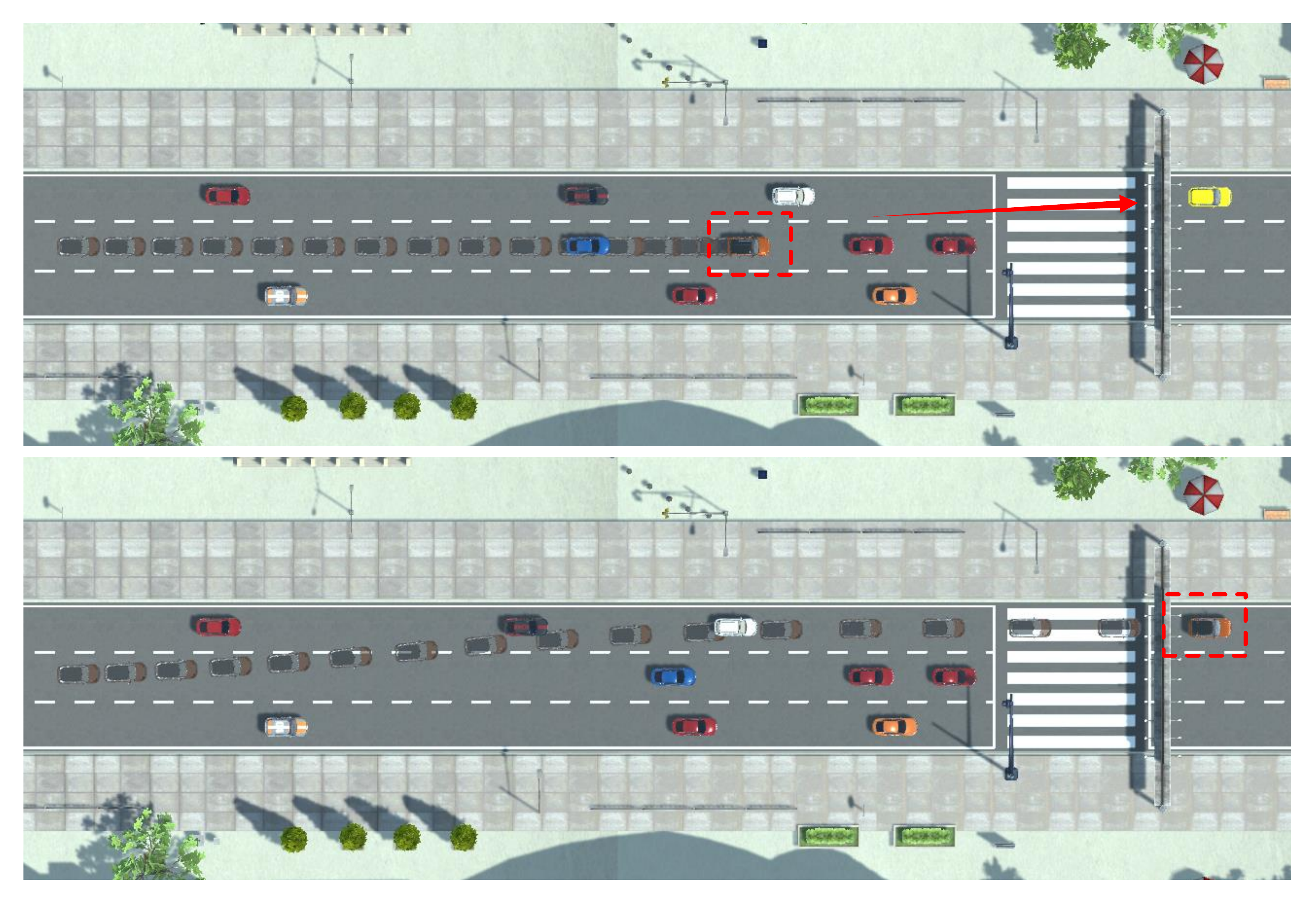}
     \caption{ The original trajectories (top) and the keyframe controlled trajectories (bottom) of running the red light in the crosswalk scenario. }
    \label{fig:casecrosswalk}
\end{figure}

\begin{figure}[t]
\setlength{\belowcaptionskip}{-0.2cm}
    \centering
    \includegraphics[width=0.99\columnwidth]{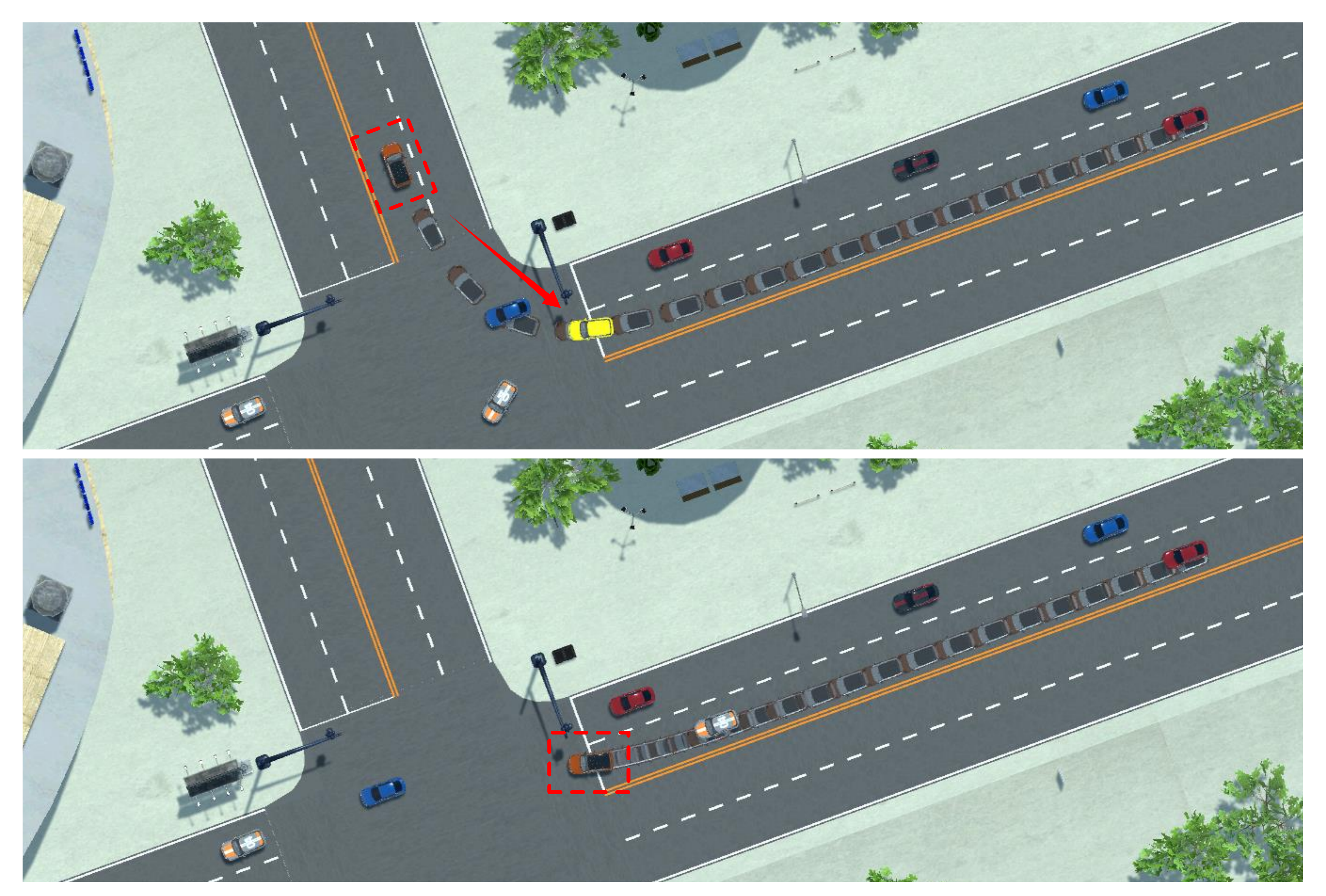}
     \caption{ The original trajectories (top) and the keyframe controlled trajectories (bottom) of yielding to the oncoming vehicles in the intersection scenario. }
    \label{fig:caseintersection}
\end{figure}

\subsection{Keyframe Controlled Trajectories}

We designed some cases to show the results of keyframe controlled simulation with our method. In the following figures, we use a yellow vehicle to represent the position constraint and a red box to frame the time constraint of the keyframe. That is to say, we want the vehicle to reach the position marked by the yellow vehicle at the moment framed by the red box. The edited vehicles are shown with its historical states, while the other vehicles are only shown with their final states.

The first case was generated in the scenario with a curvy road, which we have already shown in Fig. \ref{fig:teaser}. The specific vehicle drove along the rightmost lane originally. We set a keyframe in the center lane and assigned the vehicle to reach the position at the framed time. As a result, a new reference path through the position was planned. Then, the vehicle changed the lane and decelerated to arrive at it on time. % (See Fig. \ref{fig:casecurvy}). 

The second case was generated in the crosswalk scenario. The specific vehicle braked in the center lane and waited for the red light originally. We set a keyframe to assign the vehicle to overtake and get through the crosswalk instead of waiting. As a result, a new reference path was planned to lead the vehicle to the lane which had not been blocked yet. The vehicle accelerated to run the red light (see Fig. \ref{fig:casecrosswalk}). 

The third case was generated in the interaction scenario. The specific vehicle turned right in the left lane without yielding to the oncoming vehicles driving forward in the right lane originally. We set a keyframe to assign it to wait before turning. In this case, we only needed to specify the position and the corresponding arrival time since the reference path for the vehicle was not changed. As a result, the vehicle decelerated and yielded at the intersection to let the vehicles in the right lane go first (see Fig. \ref{fig:caseintersection}).

\subsection{Performance and Comparison}

\begin{table}[]
% first table
\begin{minipage}[!t]{\columnwidth}
\setlength{\belowcaptionskip}{0.2cm}
\renewcommand{\arraystretch}{1.2}
\centering
\setlength{\tabcolsep}{3.3mm}
{
    \begin{tabular}{|c|c|}
    \hline
    \textbf{Discretization Timestep} $\Delta{\Tilde{t}}$ & \textbf{State-Time Search Time} \\ \hline
    0.5                     & 0.173                  \\ \hline
    0.25                    & 36.924                 \\ \hline
    0.1                     & $-$                      \\ \hline
    \end{tabular}
}
\caption{The state-time search time (s) over different discretization timesteps (s). Due to memory limitation, it is hardly to obtain the state-time search time when discretization timestep is 0.1s.}
\label{tab:sttperformance}
\vspace{2mm}
\end{minipage}

% second table
\begin{minipage}[!t]{\columnwidth}
\setlength{\belowcaptionskip}{-0.2cm}
\renewcommand{\arraystretch}{1.2}
\centering
\setlength{\tabcolsep}{1.0mm}
{
    \begin{tabular}{|c|c|}
    \hline
    \textbf{Simulation Timestep} $\Delta{t}$ & \textbf{Adjoint-based Optimization Time} \\ \hline
    0.5                 & 0.002                           \\ \hline
    0.1                 & 0.004                           \\ \hline
    0.05                & 0.011                           \\ \hline
    0.01                & 0.185                           \\ \hline
    0.005               & 0.699                           \\ \hline
    \end{tabular}
}
\caption{The adjoint-based optimization time (s) over different traffic simulation timesteps (s) for a certain trajectory.}
\label{tab:adjperofmance}
\end{minipage}
\end{table}

To evaluate the performance of our coarse-to-fine optimization process, we performed a series of experiments with different timesteps or initialization for the adjoint method. The following experiments are based on a keyframe which constrains a vehicle to travel 100 meters in 10 seconds along its reference path, setting the number of the maximum iterations of the adjoint method to 100.

%\begin{figure}[t]
%    \centering
%    \includegraphics[width=0.92\columnwidth]{figures/comparison vel.pdf}
%     \caption{ The longitudinal speeds of the vehicle generated by state-time search and our method with the same keyframe constraints. }
%    \label{fig:cmpvel}
%\end{figure}

%\begin{figure}[t]
%    \centering
%    \includegraphics[width=0.92\columnwidth]{figures/comparison loss.pdf}
%     \caption{ The optimization loss over different numbers of iterations of the adjoint method initialized with average speed and our method with the same keyframe constraints. }
%    \label{fig:cmploss}
%\end{figure}

\begin{figure}[t]
\setlength{\belowcaptionskip}{-0.2cm}
    \centering
    \includegraphics[width=\columnwidth]{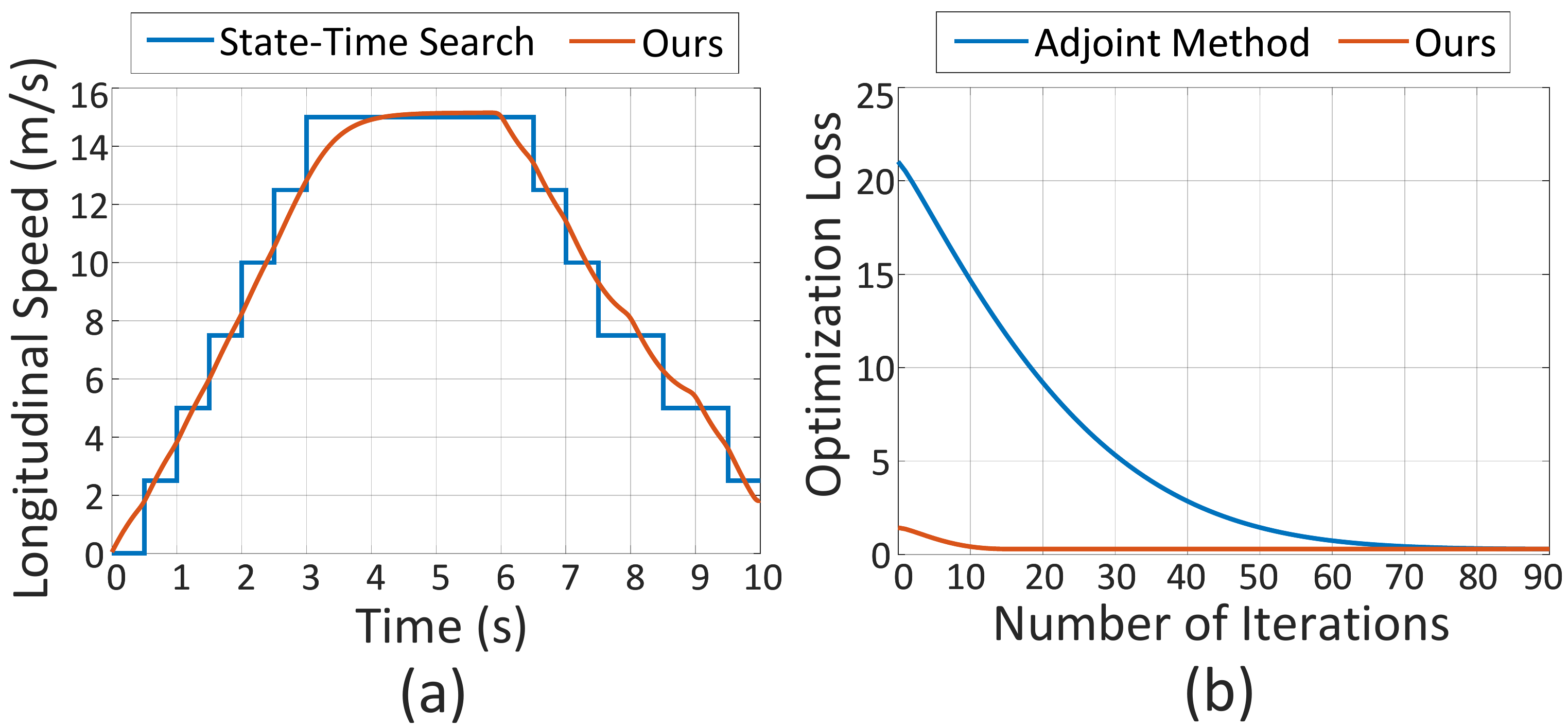}
     \caption{ (a) The longitudinal speeds of the vehicle generated by state-time search and our method with the same keyframe constraints. (b) The optimization loss over different numbers of iterations of the adjoint method initialized with average speed and our method with the same keyframe constraints.}
    \label{fig:comparisons}
\end{figure}

We searched for the coarse trajectory in the state-time space with different discretization timesteps $\Delta{\Tilde{t}}$=0.5s, 0.25s and 0.1s. The corresponding search times are shown in Table \ref{tab:sttperformance}. $\Delta{\Tilde{t}}$=0.5s was used to reconstruct the traffic flows In \cite{sewall2010virtualized}. When $\Delta{\Tilde{t}}$ gets smaller, the generated trajectories with state-time search may become smoother, but search time and required memory will also increase rapidly. According to Table \ref{tab:sttperformance}, the search time is 36.924s when $\Delta{\Tilde{t}}$=0.25s. Due to the memory limitation, the search time can hardly be obtained when $\Delta{\Tilde{t}}$=0.1s. Therefore, as mentioned earlier, it is impractical to obtain plausible keyframe controlled results by using a smaller discretization timestep in state-time space.

We optimized the trajectory using the adjoint method with different simulation timesteps $\Delta{t}$. The corresponding optimization times are shown in Table \ref{tab:adjperofmance}. The optimization time mainly depends on the number of frames of the trajectory, so it becomes more time-consuming when $\Delta{t}$ gets smaller. In our implementation, we chose $\Delta{\Tilde{t}}$=0.5s and $\Delta{t}$=0.01s in order to strike a balance between the total optimization time and the plausibility of generated trajectories. Thus the total optimization time with the given keyframe using our coarse-to-fine optimization is about 0.358s for 100 iterations.

We compare the results generated by state-time search only and our method. The longitudinal speeds along the generated trajectories are shown in Fig. \ref{fig:comparisons}~(a). The speed of the state-time search result changes suddenly since the vehicle can only choose the determined acceleration values in a discrete set at each timestep, which will cause implausibilities. In comparison, the result generated by our coarse-to-fine optimization is much smoother.

We further compare the loss during the optimization of the adjoint method and our method, and the loss values over different numbers of iterations are shown in Fig. \ref{fig:comparisons}~(b). For the adjoint method, we initialize the controlling desired speeds as the average speed that the vehicle needs to travel from the start to the key-frame position within the time. The loss achieves convergence at approximately the $70th$ iteration. In comparison, with our coarse-to-fine process, we initialize the controlling desired speeds better by utilizing the information of the coarse trajectory search in the state-time space. The loss at the beginning is thus much smaller than the optimization initialized with average speed and can achieve fast convergence at approximately the $15th$ iteration. Therefore, we can also give an early stop if the loss no longer decreases to further reduce the optimization time in our method.

Finally, the key aspects of our method in comparison with previous works are summarized in Table \ref{tab:cmpmethods}. Our method includes all the positive aspects of the previous works such as generating smooth traffic motions, allowing to simulate in complex scenarios and edit vehicles during the simulation. Furthermore, since our traffic simulation is based on the force models, we can provide spatio-temporal key frames to regulate vehicles and generate desired trajectories in a more intuitive way.

\begin{table}[]
\setlength{\belowcaptionskip}{-0.22cm}
\centering
\normalsize
\renewcommand\arraystretch{1.2}
\begin{tabular}{|c|c|c|c|c|c|}
%Method & Simulation Model & Complex Scenarios & Interactively Editing & Spatial-Temporal Keyframes \\
\hline
\textbf{Method} & \textbf{(a)} & \textbf{(b)} & \textbf{(c)} & \textbf{(d)} & \textbf{(e)} \\ \hline
\cite{sewall2010virtualized}    & State-Time Search & \ding{55} & \ding{51} & \ding{55} &\ding{51} \\ \hline
\cite{chao2021calibrated}       & Force-based & \ding{51} & \ding{55} & \ding{55} & \ding{55} \\ \hline
\cite{han2022traedits} & Data-driven & \ding{51} & \ding{51}   & \ding{51}   & \ding{55} \\ \hline
Ours                            & Force-based & \ding{51} & \ding{51}   & \ding{51}   & \ding{51}   \\ \hline
\end{tabular}
\caption{Comparison with previous methods. Several criteria are presented: (a) Traffic simulation model, (b) Smooth traffic motions, (c) Allowing to simulate in complex scenarios, (d) Allowing to interactively edit vehicles during the simulation, (e) Allowing to constrain vehicles with spatio-temporal keyframes. }
\label{tab:cmpmethods}
\end{table}

%\subsection{User Study}

\section{Discussion}
\subsection{Conclusion and Limitation}
\label{section:conclusion&limitation}

We have presented a novel traffic trajectory editing method that allows users to specify spatio-temporal keyframes to control vehicles' behaviours. We propose a force-based traffic simulation framework containing self-motivated force, path keeping force and collision avoidance force. It mainly updates vehicles based on the Frenet coordinates along vehicles' reference paths which can be defined manually. To provide keyframe controls, we propose a coarse-to-fine optimization process. First, we discretize the state-time space along the path, construct a state-time graph and plan a coarse trajectory from the start to the keyframe node. Then, we utilize the coarse trajectory to initialize the adjoint method and efficiently find a set of optimal controlling desired speeds to generate a finer trajectory based on our force-based simulation.

Though the proposed method is promising, it still has some limitations. Firstly, the keyframe constraints may become inoperative if the environment is congested. As we stated in Section \ref{subsection:sttsearch}, the state-time search regards the neighbors of an individual as static obstacles, which means its neighbors will not be optimized at the same time if we use keyframes to constrain it. If the vehicle is blocked when it tries to meet the keyframes during the state-time search, the result will be replaced by a trajectory that can reach the node closest to the actual goal. Though the problem can be solved by iteratively editing the surrounding vehicles who block the individual until it can meet the keyframes, we believe that a optimization process taking all the possible vehicles into consideration at once will be a worthwhile endeavor. Secondly, the simulation results may become irregular due to users' arbitrary edits. For example, a vehicle may decelerate for safety and comfort when it follows a path with sharp curves rather than maintaining a high speed in the real traffic. This problem can also be solved by interactively setting keyframes or editing vehicle's desired speed by users to make the behaviours more perceptually realistic. 

\subsection{Failure Cases and Refinements}
\label{section:failures}

%\textcolor{red}{We give some failure cases related to the mentioned-above limitations, attached with the refinements by manually providing more keyframes to demonstrate the capability of our framework. For more details, please refer to our supplementary video.}
To demonstrate our framework's capability, even for some failure cases caused by the aforementioned limitations, our method is still capable of refining them by manually providing more keyframes. For more information, please see our supplementary video.

\begin{figure*}[ht]
    \centering
    \includegraphics[width=0.89\textwidth]{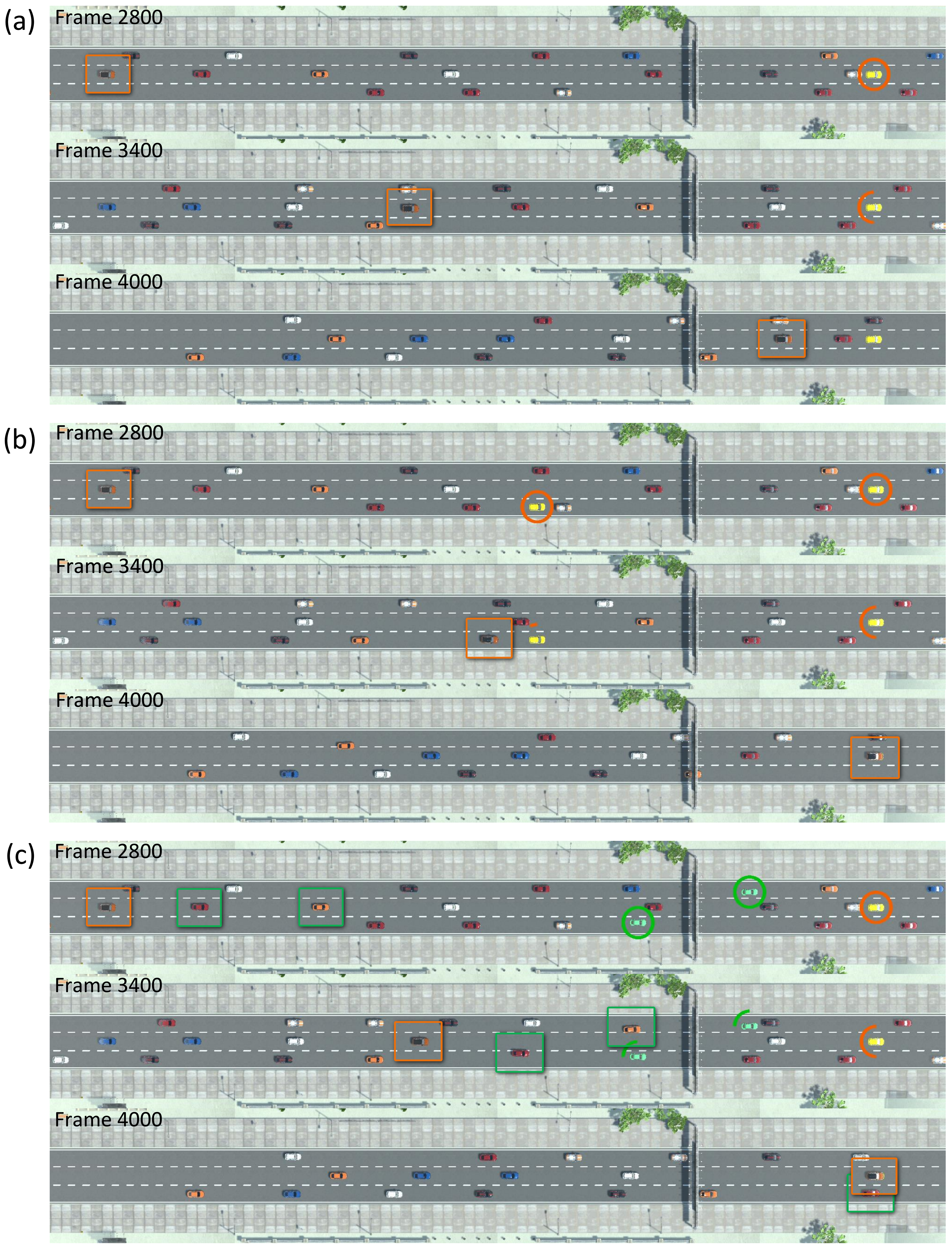}
        \caption{The first failure case. (a) The vehicle misses the keyframe due to traffic congestion. (b) The first solution is to give the vehicle another keyframe that causes it to overtake the leader. (c) The second solution is to assign keyframes to its leaders, forcing them to yield. }
    \label{fig:failure1}
%\vspace{-0.45cm}
\end{figure*}

%\textcolor{red}{The first case is the vehicle fails to meet the keyframe due to the congestion (See Fig.~\ref{fig:failure1}~(a)). We provide two ways to solve the problem. The first solution is assigning another keyframe to the vehicle which makes it overtake the leader (See Fig.~\ref{fig:failure1}~(b)). The second solution is assigning keyframes to its leaders which make them give way to it (See Fig.~\ref{fig:failure1}~(c)). As a result, both of the refinements succeed in making the vehicle meet the keyframe eventually.}
The first scenario is that the vehicle misses the keyframe due to traffic congestion (see Fig.~\ref{fig:failure1}~(a)). We offer two solutions to the problem. The first solution is to give the vehicle another keyframe that causes it to overtake the leader (see Fig.~\ref{fig:failure1}~(b)). The second solution is to assign keyframes to its leaders, forcing them to yield (see Fig.~\ref{fig:failure1}~(c)). As a result, both refinements eventually succeed in getting the vehicle to meet the keyframe.

\begin{figure*}[ht]
    \centering
    \includegraphics[width=0.7\textwidth]{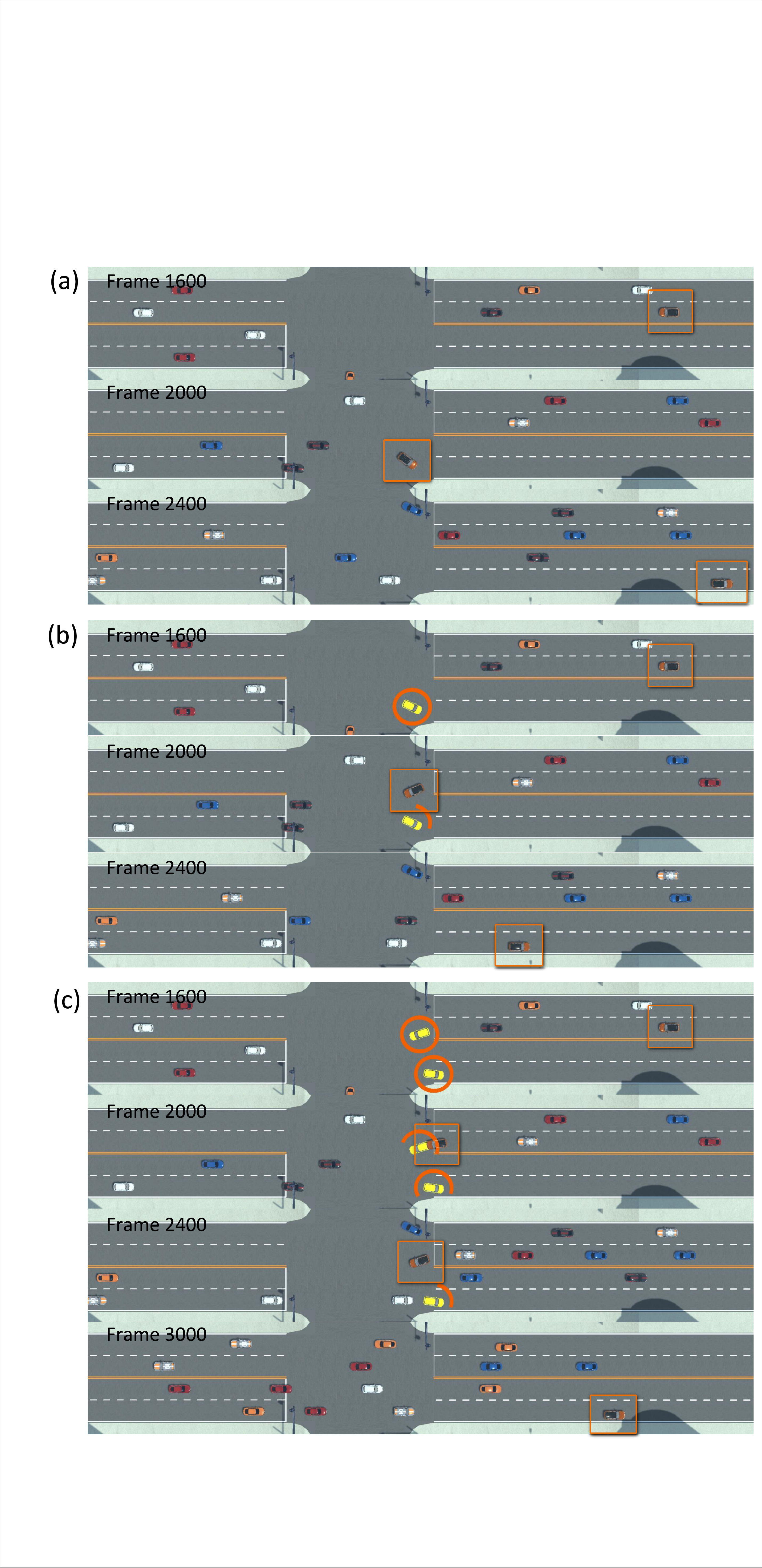}
        \caption{The second failure case. (a) The vehicle behaves abnormally when performing a high-speed U-turn. (b) The first solution is to assign a keyframe that causes the vehicle to simply decelerate when passing through the curve. (c) The second solution is to assign two keyframes, one to wait for oncoming vehicles and one to complete the U-turn, to further prevent aggressive driving and maintain polite behavior.}
    \label{fig:failure2}
%\vspace{-0.45cm}
\end{figure*}

%\textcolor{red}{The second case is the vehicle behaves irregularly when it takes a U-turn at a high speed (See Fig.~\ref{fig:failure2}~(a)), since in the real-world traffic, drivers usually tend to slow down for safety and comfort when following a sharper path. We also provide two ways to make the behaviour more perceptually realistic. The first solution is assigning a keyframe to make the vehicle simply decelerate when crossing the curve (See Fig.~\ref{fig:failure2}~(b)). The second solution is assigning two keyframes to make it wait for the oncoming vehicles at first and then finish the U-turn, to further prevent aggressive driving and keep polite (See Fig.~\ref{fig:failure2}~(c)).}
The second scenario is that the vehicle behaves abnormally when performing a high-speed U-turn (see Fig.~\ref{fig:failure2}~(a)). This is due to the fact that in real-world traffic, drivers tend to slow down for safety and comfort when following a more curved path. We also provide two methods for improving the behavior's perceptual realism. The first solution is to assign a keyframe that causes the vehicle to simply decelerate when passing through the curve (see Fig.~\ref{fig:failure2}~(b)). The second solution is to assign two keyframes, one to wait for oncoming vehicles and one to complete the U-turn, to further prevent aggressive driving and maintain polite behavior (see Fig.~\ref{fig:failure2}~(c)).

\section*{Acknowledgment}

%The preferred spelling of the word ``acknowledgment'' in America is without an ``e'' after the ``g''. Avoid the stilted expression ``one of us (R. B. G.) thanks $\ldots$''. Instead, try ``R. B. G. thanks$\ldots$''. Put sponsor acknowledgments in the unnumbered footnote on the first page.

Xiaogang Jin was supported by the National Natural Science Foundation of China (Grant No. 62036010) and the Key Research and Development Program of Zhejiang Province (Grant No. 2020C03096).

%\section{Reference}

% bibtex
\bibliographystyle{eg-alpha-doi} 
\nocite{*}
%\bibliography{egbibsample}       
\bibliography{egbib}

% biblatex with biber
% \printbibliography                

% \input{template}

\end{document}